\documentclass[english]{article}
\usepackage[latin9]{inputenc}

\makeatletter

\providecommand{\tabularnewline}{\\}

\makeatother

\usepackage{babel}
\begin{document}

\title{On the Coulomb Interaction in a Nucleus for Odd J States}

\author{Larry Zamick \\
 Department of Physics and Astronomy, \\
 Rutgers University, Piscataway, New Jersey 08854}

\maketitle

\section{Abstract}

With a certain approximation for the Coulomb matrix elements in a
single j shell of protons and neutrons it is found that wave functions
of states of odd angular momentum J in an even-even nucleus are not
strongly affected by their presence,

\section{Introduction}

We make a simple estimate of the matrix elements of the Coulomb interaction
and then calculate the isospin mixing of the lowest J=2$^{+}$ T=0
and lowest J=2$^{+}$ T=1 states in $^{44}$Ti.

\section{The Calculation.}

If one examines the Coulomb 2 particle matrix elements in a single
j shell one notices that the main affect of thir presence is to add
a repulvive term te the J=0 matrix element, i.e. like anti-pairing.What
are the conquenses of assuming this simple model?

Let us start with 2 protons and 2 neutrons in a sngle j shell. The
wave functions can be written as follows:

$\psi^{\alpha}$$^{J}$= $\Sigma$ D$^{\alpha J}$(J$_{p}$J$_{n}$)
{[} (jj)$^{J_{p}}$(jj)$^{J_{n}}${]}$^{J}$

where $\Sigma$ D$^{\alpha J}$(J$_{p}$J$_{n}$) is the probability
amplitude that the 2 protons couple to J$_{p}$ and the 2 neutrons
to J$_{n}$. 

To satisfy the Pauli principle J$_{p}$ must be even and likewise
J$_{n}$. Odd J (T=0) states can only exist in the proton-neutron
channel.

Note that for odd J it is easy to see that J$_{p}$ cannot be equal
to sero. If it were zero then J would equal J$_{n}$. But J$_{n}$
is even.Hence the odd J wave function is insensitive to the 2 particle
J=0 Coulomb matrix element, which ,in our approximation is the only
non-vanishing Coulomb matrix element.

Let us contrast the above with what happens if we add a consant to
all J=0 matrix elements- pp,nn and np. In that case there would be
contribuiotns from the pp channel, the nn channel and the np channel.
For the pp and nn channels the expression is easy Concerning the latter
the expression for a matrix element os a 2p-2n system is

The expresion is \textless{} (J$_{p^{'}}$J$_{n^{'}}$)$^{J}$\textbar{}
V\textbar{} (J$_{p}$J$_{n}$)$^{J}$\textgreater{} = $\delta_{J_{p^{'}}}$$_{J_{p}}$
$\delta$$_{J_{n^{'}}}$$_{J_{n}}$((E(J$_{p}$) + E(J$_{n}$))

+ 4$\Sigma$((jj)$^{J_{p^{'}}}$(jj)$^{J_{n^{'}}}$\textbar{} (jj)$^{J_{A}}$(jj)$^{J_{B}}$)$^{J}$
$\Sigma$((jj)$^{J_{p}}$(jj)$^{J_{n}}$\textbar{} (jj)$^{J_{A}}$(jj)$^{J_{B}}$)$^{J}$
\textless{} (jj)$^{J_{B}}$\textbar{} V$_{np}$\textbar{} (jj)$^{J_{B}}$\textgreater{}

Although J$_{B}$ has to be even J$_{A}$ $ $can be even or odd.
Hence we can have a contribution where 

J$_{B}$ is equal to zero and J$_{A}$is equal to J. However in the
Coulomb case we do not have this neutron -proton contribution.

With this approximation of the Coulomb matrix elements there is also
no mixing for states of even J with high spin. If the highest J for
2 identical nucleons is J$_{m}$then states of 2 protons and 2 neutrons
with J greater then J$_{m}$will not have any components with J$_{P}$equal
to zero. For example, in the f$_{7/2}$shell the maximum J for 2 protons
is 6 and so states with J= 8,10 and12 will not have components with
J$_{P}$ equal to zero.

.

.For an even J less than J$_{m}$ the coupling matrix element between
a T=0 and T=1 state in this approximation is 

C D$^{T=0}$ (0, J) D$^{T=1}$ (0,J) where C is the strength of the
Coulomb interaction for 2 protons on a state with J$_{P}$ equal to
zero.

As an illustration we show the lowest J=2 T=0 and J=2 T=1 states as
calculated with the MBZE wave functions {[}1{]}.. Note that the MBZE
2 body matrix elemtns are different from those in earlier works{[}2,3{]}.
This is especially true for the T=0 matrix 2 body elements.

\begin{minipage}[t]{1\columnwidth}%
Table I Wave functions of J=2$^{+}$ states in $^{44}$Ti with the
MBZE interaction-f$_{7/2}$shell. E(T=0) 1.1631 MeV, E(T=1)=5.2366
MeV.%
\end{minipage}

.

.%
\begin{tabular}{|c|c|c|}
\hline 
J$_{P}$,J$_{N}$ & D$^{T=0}$(J$_{P}$J$_{N}$) & D$^{T=1}$(J$_{P}$J$_{N}$)\tabularnewline
\hline 
\hline 
0,2 & 0.6099 & -0.6370\tabularnewline
\hline 
2,0 & 0.60 99 & 0.6370\tabularnewline
\hline 
2,2 & -0.3538 & 0\tabularnewline
\hline 
2,4 & 0.2416 & -0.2962\tabularnewline
\hline 
4,2 & 0.2416 & 0.2932\tabularnewline
\hline 
4,4 & -0.0591 & 0\tabularnewline
\hline 
4,6 & 0.0613 & -0.0909\tabularnewline
\hline 
6,4 & 0.0613 & 0.0909\tabularnewline
\hline 
6,6 & 0.0563 & 0\tabularnewline
\hline 
\end{tabular}

. .

Using the approximate formula the mixing amplitude between these 2
J=2$^{+}$states is C 0.6099{*} -0.6370/(1.1631-5.2366) = 0.09537
C. Taking a resonable estimate C=1.0 MeV we find the probability of
a T=1 admixture in a basically J=2$^{+}$ T=0 state is 0.0091.

\end{document}